\documentclass[prd,aps,floats,twocolumn,nofootinbib,superscriptaddress]{revtex4}
\usepackage{commath}
\pdfoutput=1
\usepackage[usenames, dvipsnames]{color}
\usepackage{graphicx, array}
\usepackage{multirow}
\usepackage{graphics}
\usepackage{graphics,epsfig}
\usepackage{amssymb}
\usepackage{amsmath}
\usepackage{ulem}
\usepackage{multirow}
\usepackage{subfigure}
\usepackage{bm}
\usepackage{txfonts}
\usepackage{cancel}
\usepackage{url}
\usepackage{hyperref}
\usepackage{lipsum}
\usepackage{graphicx}
\usepackage{soul}

\newcolumntype{C}[1]{>{\centering\arraybackslash}m{#1}}

\def\be{\begin{equation}}
\def\ee{\end{equation}}
\def\bi{\begin{itemize}}
\def\ei{\end{itemize}}
\def\ben{\begin{enumerate}}
\def\een{\end{enumerate}}
\def\bt{\begin{tabular}}
\def\et{\end{tabular}}
\def\bc{\begin{center}}
\def\ec{\end{center}}

\def\bea{\begin{eqnarray}}
\def\eea{\end{eqnarray}}

\def\ba{\begin{eqnarray}}
\def\ea{\end{eqnarray}}

\let\oldhat\hat

\renewcommand{\hat}[1]{\oldhat{\boldsymbol{\mathbf{#1}}}}

\graphicspath{ {./images/},{ ./images/ratioFigs/},{ ./images/windowFigs/},{ ./Version1/}}

\begin{document}
\input{epsf}

\title{kSZ Pairwise Velocity Reconstruction with Machine Learning}

\author{Yulin Gong}
\author{Rachel Bean}
\affiliation{Department of Astronomy, Cornell University, Ithaca, NY 14853, USA}

\begin{abstract} 
We demonstrate that pairwise peculiar velocity correlations for galaxy clusters can be directly reconstructed from the kinematic Sunyaev-Zel'dovich (kSZ) signature imprinted in the CMB using a machine learning model with a gradient boosting algorithm trained on high-fidelity kSZ simulations. The machine learning model is trained using six to seven cluster features that are directly related to observables from CMB and large-scale structure surveys. We validate the capabilities of the approach in light of the presence of primary CMB, detector noise, and potential uncertainties in the cluster mass estimate and cluster center location. The pairwise velocity statistics extracted using the techniques developed here have the potential to elicit valuable cosmological constraints on dark energy, modified gravity models, and massive neutrinos with kSZ measurements from upcoming CMB surveys, including the Simons Observatory, CMB-S4 and CCAT, and the DESI and SDSS galaxy surveys. 
\end{abstract}
\maketitle

\section{Introduction}
\label{sec:intro}
The origin of accelerated cosmic expansion remains a critical outstanding problem in physics.  Measurements of the cosmic microwave background (CMB) radiation \cite{Planck:2013pxb, Planck:2015mrs, Planck:2018vyg, WMAP:2003ivt, WMAP:2010qai, WMAP:2012nax, Calabrese:2013jyk, Story:2012wx}, Baryon acoustic oscillations (BAO)  (e.g. \cite{SDSS:2005xqv, Percival:2007yw, SDSS:2009ocz, Padmanabhan:2012hf, Kazin:2014qga, eBOSS:2020yzd}) and type 1a supernovae (e.g. \cite{SupernovaSearchTeam:1998fmf, SupernovaCosmologyProject:1998vns, T:2015dqj, Zhang:2017aqn}) together provide exquisite constraints on the expansion history of the universe. This expansion history is consistent with the standard cosmological model, which assumes General Relativity (GR) and a cosmological constant, $\Lambda$, the simplest form of dark energy, as the component of the cosmic matter density proposed to explain the accelerated expansion of the universe  
(e.g. \cite{Peebles:2002gy, Uzan:2010ri, Jain:2013wgs, Joyce:2014kja}). Given the fine-tuning and coincidence problems \citep{Weinberg:1988cp, Egan:2007ht, Quartin:2008px, Shaw:2010pq, Velten:2014nra, DelPopolo:2016emo, Cunillera:2021izz, Shimon:2022ehj, Jalalzadeh:2022dlj} related to the discordance between the observed value of $\Lambda$ and those naturally predicted from theory, modifications of gravity, beyond GR, have also been actively considered as alternative explanations for the accelerated expansion (e.g. see \citep{Clifton:2011jh} for a review). Such modifications can be developed to match a $\Lambda$CDM expansion history but concurrently predict differences in the growth and dynamical properties of inhomogeneities, probed through the clustering and dynamical properties of large-scale structure (LSS), galaxies and clusters of galaxies (e.g. \citep{Clifton:2011jh, Bellini:2014fua, deMartino:2015zsa, Cusin:2017wjg, LHuillier:2017pdi, Perenon:2019dpc, Jiang:2023nzz}).   

In the context of this paper, we focus on the use of the dynamics of galaxy clusters as a cosmological tracer of the underlying gravitational field.  Galaxy clusters can be detected observationally through unique signatures left in CMB photons when they interact with the hot gas of a galaxy cluster, the Sunyaev-Zel’dovich (SZ) effect. The SZ effect occurs when CMB photons interact with electrons in galaxy clusters and can be separated into two principle components:  the thermal Sunyaev-Zel’dovich effect (tSZ) and the kinematic Sunyaev-Zel’dovich effect (kSZ) \cite{Sunyaev:1970eu, Sunyaev:1972eq, Sunyaev:1980nv} (and see reviews \cite{Birkinshaw:1998qp, Carlstrom:2002na}). The tSZ is caused by the hot electrons with random velocities boosting the blackbody spectrum imprinting a characteristic frequency dependent-signature that facilitates its isolation from the CMB through multi-frequency measurements. The kSZ is produced by the peculiar (bulk) line of sight motion of a galaxy cluster creating a Doppler shift of the CMB spectrum, and is one order of magnitude smaller than the tSZ effect and largely frequency-independent making it harder to extract. 

The kSZ signature is an observational tracer of the underlying peculiar velocities of clusters and, in turn, the gravitational potential \cite{Kosowsky:2009nc, Mitchell:2020fnj, Okumura:2021xgc, Hou:2023kfp}. The gravitational attraction between pairs of clusters creates inherent in-fall towards each other. This gravitational attraction leads to a pairwise correlation statistic that can provide a potentially sensitive measurement of the large-scale velocity field \citep{1980lssu.book.....P, Diaferio:1999ig, Ferreira:1998id, Aghanim:2001yu}. 

Despite the comparative challenges in kSZ versus tSZ detection, the prospect of extracting the peculiar velocities of galaxy clusters from the kSZ effect is actively investigated as it offers a potentially powerful technique for cosmological inference (e.g.\cite{ Hernandez-Monteagudo:2005xtx, DeDeo:2005yr, Bhattacharya:2006ke, Bhattacharya:2007sk, Fosalba:2007bx,  Planck:2013rgv, Soergel:2017ahb, Planck:2017xaj}),  and complementary constraints on the properties of gravity on cosmic scales and the neutrino mass sum to those from galaxy lensing and clustering measurements \citep{ Mueller:2014dba, Mueller:2014nsa, Kuruvilla:2020gcm, Zheng:2020qcw}.

The pairwise cluster momentum has been a principal estimator used to extract out the kSZ signal. The mean pairwise cluster momentum estimator was first detected from the kSZ signal in \citep{Hand:2012ui} with the Atacama Cosmology Telescope (ACT) CMB observations \citep{Swetz:2010fy} and the LSS surveys of Sloan Digital Sky Survey (SDSS) \citep{SDSS:1998xob, SDSS:2006srq}. Measurements from subsequent ACT and SDSS data releases, covering a greater survey area and expanded catalogs, have also been made \citep{DeBernardis:2016pdv, Calafut:2021wkx}. The kSZ signal has also been detected using the same estimator, by the Planck collaboration with SDSS \cite{Planck:2015ywj}  and the South Pole Telescope collaboration (SPT) \cite{Carlstrom:2009um} using a cluster catalog from the Dark
Energy Survey (DES) \citep{DES:2016umt, SPT-3G:2022zrq}.

The kSZ signal can also be utilized with techniques other than pairwise momentum, including projected fields \citep{Hill:2016dta, Ferraro:2016ymw, Kusiak:2021hai, Bolliet:2022pze, Patki:2023fjz}, velocity reconstruction \citep{Li:2014mja, ACTPol:2015teu, Tanimura:2022fde, DES:2023mug},  kSZ tomography \citep{Smith:2018bpn, Sato-Polito:2020cil}, individual cluster measurement \citep{Sayers:2013ona, Mittal:2017hwf}, CMB temperature dispersion \citep{Planck:2017xaj}, 21cm intensity mapping \citep{Li:2018izh}, signatures from CMB anisotropies \citep{George:2014oba}, and Fourier space analysis \citep{Li:2024svf}. 

Typically, the extraction of pairwise velocity statistics is undertaken by combining the pairwise kSZ momentum estimators with observational estimation of the optical depths for the same cluster sample. This requires a clear understanding of how the kSZ signal and optical depth measurements are related in terms of their respective sampling of the cluster properties \cite{Battaglia:2016xbi, Shaw:2010mn}.  Optical depth estimates can be obtained using one or more observables such as the tSZ effect and X-ray observations, however, the optical depth estimation presents challenges in introducing additional uncertainties and potential biases \cite{Dolag:2013hj, Flender:2016cjy, Mittal:2017hwf, Calafut:2021wkx, Vavagiakis:2021ilq, SPT-3G:2022zrq, Gong:2023hse}. 

In this work, we demonstrate the efficacy of an alternative approach focused on extracting cluster velocity information from the kSZ measurements directly, using a machine learning technique. These velocity estimates are then used to infer the pairwise velocity,  rather than calculating it from separate measurements of the pairwise momentum and optical depth.

Machine learning, a subfield of artificial intelligence and computer science, is a powerful tool that uses statistical techniques to efficiently analyze and identify patterns in massive and complex datasets to acquire knowledge that may be computationally intractable using other approaches. With the growth in complexity and volume of astronomical data,  machine learning methods have been used for a range of applications, including cluster mass estimation \citep{Green:2019uup, KodiRamanah:2020tlp, Yan:2020wsr, Ho:2019zap, Ho:2020lzz, Gupta:2020yvd, deAndres:2021tjl,  Ferragamo:2022fdr, deAndres:2022mox, Ho:2023dgt}, cluster scaling relations \citep{Shao:2021qoa, Delgado:2021cuw, Wadekar:2022bss, Wadekar:2022cyw}, simulation and maps \citep{Rodriguez:2018mjb, Mishra:2019sep, CAMELS:2020cof, Rothschild:2021ruc}, cosmological parameter constraint and estimation \citep{Pan:2019vky, Ravanbakhsh:2017bbi, Lazanu:2021tdl, Bengaly:2022cgs, Perez:2022nlv}, cosmic structure formation \citep{Lucie-Smith:2018smo, He:2018ggn, Lucie-Smith:2020orp, Petulante:2022dqc}, reconstruction of the cosmological density and velocity fields \citep{Wu:2021jsy, Wu:2023wmj, Qin:2023dew, Wang:2023hgm}, strong gravitational lensing \citep{Petrillo:2017njm, PerreaultLevasseur:2017ltk, Hezaveh:2017sht, Morningstar:2018ase}, weak gravitational lensing \citep{Tewes:2018she, Gupta:2018eev, Springer:2018aak, Ribli:2019wtw, Fluri:2019qtp, Jeffrey:2019fag, Zhang:2023pwh}, amongst others (e.g. see \citep{baron2019machine, Mehta_2019, Carleo:2019ptp, Ntampaka:2019udw}). Algorithms are normally designed without specific programming of physics, hence, machine learning provides an alternative data-driven method to the physics model-driven analysis paradigm. 

Deep learning techniques have been used to recover galaxy cluster peculiar velocities \cite{ Wang:2020kvd, Tanimura:2022fde} using 2D SZ images and 3D galaxy distributions respectively with a neural network. In this work, we use an alternative machine-learning technique, the Gradient Boosting algorithm that has successfully been applied to other areas of astronomy (\citep{Darya_2023, Sahakyan:2022uvr, Tolamatti:2023hef, Coronado-Blazquez:2023bbu}), to reconstruct the pairwise peculiar velocity using 1D features that are commonly distilled from observations for use in kSZ analyses. This includes the disk kSZ temperatures, redshifts, and halo mass estimates. We test the capability of the model under conditions including realistic primordial CMB and detector noise for upcoming CMB observations as well as the impact of some key potential systematic effects. We train our machine learning algorithm on one kSZ simulation and apply it to test samples from two simulation datasets that consider similar astrophysical effects and cosmological models. Both datasets provide us with complete halo information including the redshift, mass, peculiar line-of-sight velocity, kSZ temperatures, and other key properties. This information allows us to efficiently train our model to reconstruct an unbiased cluster's peculiar velocity. 

The following is a summary of the paper's outline: The theory for the kSZ signal extraction techniques and pairwise estimator is introduced in section \ref{sec:formalism}, along with a description of the kSZ simulation datasets and halo catalogs that we used in this work. In section \ref{sec:FE_LM}, we describe our feature engineering process, the structure of our machine-learning model, and covariance estimates. In section \ref{sec:results}, we present the findings, including the sensitivity to various modeling assumptions and to systematic errors that might arise from cluster center mislocation and scattering in the cluster mass estimation. We cross-validate the model by training on one set of simulations and inferring the velocities from completely distinct kSZ and halo simulations. In section \ref{sec:conc}, we conclude with a summary of the approach, the key findings, and implications for future study.

\section{Background}
\label{sec:formalism}
The primary goal is to recover an estimate of the pairwise peculiar velocity correlations for a cosmological galaxy cluster sample using the information from galaxy surveys and kSZ temperatures measured from CMB observations. In \ref{sec:data}, we discuss the simulated galaxy and CMB datasets used in this work. In section~\ref{sec:filter}, we describe the kSZ signal extraction techniques and the pairwise statistical estimator is described in section~\ref{sec:Vhat}.

\subsection{Datasets}
\label{sec:data}
In this work, we use the simulated kSZ maps from \cite{Flender:2015btu}\footnote{\url{https://www.hep.anl.gov/cosmology/ksz.html}}(the Flender simulation) to train and test our machine-learning model. The kSZ maps were generated to represent the signal expected from the passage through dark matter large-scale structures in a single realization from the Mira Universe simulation suite \citep{Heitmann:2015xma} using the HACC (Hardware/Hybrid Accelerated Cosmology Code N-body simulation) framework \citep{Habib:2014uxa}. The simulation adopts WMAP7 cosmological parameters \cite{Larson:2010gs} and is simulated with $3200^3$ particles in a $(2.1 Gpc)^3$ volume and a mass resolution of $10^{10} M_{\odot}$. A friends-of-friends algorithm is used to identify dark matter halos. The simulation provides a full-sky halo catalog spanning redshifts $0<z<1$ and masses $10^{10}M_{\odot}<M<3 \times 10^{15}M_{\odot}$. 

The kSZ signal is simulated under three different models of the intra-cluster gas: Model 1 assumes baryons trace the dark matter at all scales (i.e. gas traces mass) and includes the diffuse (non-halo) component. Model 2 follows an intra-cluster gas prescription from \citep{Shaw:2010mn} and ignores the diffuse kSZ component. Model 3 (herein FL3) is a combination of Model 2 and the diffuse kSZ components from Model 1.  As Model 3 is the most realistic kSZ model, it is the principal kSZ signal used in this work.  The Flender kSZ maps are simulated at a resolution of 0.43$^\prime$ which is comparable to the recent CMB experiments such as the ACT data \citep{Naess:2020wgi}. 

We also consider a second simulated kSZ map from \citep{Stein:2020its}\footnote{\url{https://mocks.cita.utoronto.ca/index.php/WebSky_Extragalactic_CMB_Mocks}}(the Websky simulation) to cross-validate the model that is trained on the Flender simulation. The kSZ maps in this case were generated from a lightcone realization of N-body with $6144^3$ particles in a $(7.7 Gpc)^3$ volume that covers redshift $0 < z < 4.6$ over the full-sky. The astrophysical effects and cosmological model considered in Websky are very similar to the Flender simulation such as star formation, feedback, and non-thermal pressure simulated using parametric models informed by hydrodynamical simulations. A primary distinction between the two simulations is that the resolution of the Websky simulation, 0.87$^\prime$, is lower than the Flender simulation.  

We also test the feasibility of our model to anticipate CMB observations by simulating the detector noise and primordial CMB. We run CAMB \citep{2011ascl.soft02026L} to generate the angular power spectrum of the primary CMB anisotropies with best fit Planck cosmological parameters for a zero-curvature universe \citep{Planck:2015fie}. The instrument noise is simulated in accordance with the upcoming observation from Simons Observatory (SO) \cite{SimonsObservatory:2018koc} at 145 GHz. Specifically, we simulate the instrument noise as white noise at a 6.3 $\mu K$-arcmin noise level. A Gaussian beam (FWHM = 1.4 arcmin) is also considered in accordance with the SO instrument beam at 145 GHz.   

\subsection{kSZ Effect and Signal Extraction}
\label{sec:filter}
The CMB temperature change resulting from the kSZ effect is given by: 
\begin{equation}
\label{eq:kSZ}
    \frac{\delta T_{kSZ}}{T_0} = - \int_{los} \sigma_{T}\,n_e\,\frac{v_{los}}{c}\, dl,
\end{equation}
where $\sigma_T$ is the Thomson cross-section, $n_e$ is the electron number density, c is the speed of light, $v_{los}$ is the line-of-sight peculiar velocity, and $T_0$ = 2.726K is the average CMB temperature. 

\subsubsection{Aperture Photometry}
\label{sec:ap}
Aperture photometry is a filtering method that measures an average disk temperature by averaging the temperatures within the central disk and subtracting the average temperature in a surrounding ring annulus of equal area to mitigate potential background contamination. For a disk located at position $\hat{n}$, the aperture photometry temperature can be written as
\begin{equation}
    T_{AP}(\hat{n},\theta_{AP}) = T_{disk}(\hat{n},\theta_{AP}) - T_{ring}(\hat{n},\theta_{AP}),  
\end{equation}
where angular radius $\theta_{AP}$ is the filter scale. Note that a correction factor that takes into consideration the kSZ signal in the annulus being subtracted is necessary for aperture photometry to relate $T_{AP}$ to the disk kSZ temperature \citep{Gong:2023hse}.

\subsubsection{Matched Filter}
\label{sec:mf}
Given a pre-defined template profile embedded under a noisy signal, a matched filter can be constructed to detect the presence of the template profile with minimum variance. For the kSZ signal template, we use the projected Navarro-Frenk-White (NFW) profile \citep{Bartelmann:1996hq} $\tau$ which is written as:
\begin{equation}\label{eq:nfw}
    \tau(x) = \frac{A}{x^2-1} 
    \begin{cases}
    1 - \frac{2}{\sqrt{1-x^2}}\,\text{tanh}^{-1}\sqrt{\frac{1-x}{x+1}} & 0 < x < 1 \\
    0 & x = 1 \\
    1 - \frac{2}{\sqrt{x^2-1}}\,\text{tan}^{-1}\sqrt{\frac{x-1}{x+1}} & x > 1,
    \end{cases}
\end{equation}
where $x = \theta/\theta_{s}$, and $\theta_{s}$ is the scale angle. In Fourier space, the matched filter can be written as:
\begin{equation}\label{eq:mf}
    \Psi(k) = \sigma^2 \frac{\tau(k)B(k)}{P(k)},
\end{equation}
where $\tau(k)$ is the signal template profile in Fourier space, $B(k)$ is the instrument beam, and $\sigma^2$ denotes the filter variance that:  
\begin{eqnarray}\label{eq:sigma}
    \sigma^2 = \left[\int \frac{|\tau(k)B(k)|^2}{P(k)} \frac{d^2k}{(2\pi)^2}\right]^{-1}.
\end{eqnarray}
where $P(k)$ is the noise power spectrum. 

We implement the matched filter in Fourier space by filtering a postage cutout centered at the cluster, and the matched filtered temperature, $T_{MF}(\hat{n},\theta_s)$, is measured from the filtered cutout. 

Matched filters are commonly used to determine the central peak amplitude, however in this work, we implement the matched filter differently. Following \citep{Gong:2023hse}, a matching filter, as opposed to aperture photometry, can directly extract an unbiased disk kSZ temperatures using a carefully calibrated NFW signal template profile parameterized solely with the scale radius $\theta_s$. Specifically, the matched filter is built in a way similar to minimizing the mean sum of squares (MSS) between the difference of the matched filtered amplitude and the true amplitude of the kSZ radial profile at each position $r$ within a certain angular scale $\theta$, denoted as:
\begin{equation}\label{eq:MF_MSE}
    MSS = \frac{1}{N}\sum_{r=0'}^{r=\theta'}(T_r - T_{MF,r})^2,
\end{equation}
where $T_{r}$ is the signal amplitude at $r$. This is done by tuning the filter's signal temple profile $\tau$ through $\theta_s$. In this work, $\theta_s$ is set by selecting a subsample of 20,000 halos from the sample at random and searching for the value of $\theta_s$ that minimizes the difference of the average disk temperature between the true and matched filtered estimate for this subsample.

\subsection{Pairwise estimator}
\label{sec:Vhat}

The pairwise velocity of clusters given each of their line of sight peculiar velocity $v_{i}$, is given by \citep{Ferreira:1998id}:
\begin{equation}\label{eq:Vhat}
    \hat{V} (r)= -\frac{\sum_{ij}(v_i - v_j)c_{ij}}{\sum_{ij}c_{ij}^2},
\end{equation}
where the sum is over all cluster pairs with separation $r_{ij} = |r_i - r_j|$ that fall in a radial bin centered around distance $r$. $c_{ij}$ is the geometric weight projected along the line of sight given by,
\begin{equation}
    c_{ij}=\hat{r}_{ij} \, \dot{}\,\frac{\hat{r_i}-\hat{r_j}}{2}=\frac{(r_i-r_j)(1+cos \, \theta)}{2\sqrt{r_i^2\,+r_j^2\,-2 r_i r_j cos \, \theta}},
\end{equation}
where $r_i$, $r_j$ are the comoving distance, and $\theta$ is the angle between vectors $r_i$ and $r_j$. 

Replacing the velocity with the kSZ temperature of each cluster gives the pairwise kSZ momentums estimator and is written as \citep{Hand:2012ui}:
\begin{equation}\label{eq:pob}
    \hat{p} (r_i,z_i)= -\frac{\sum_{ij}(\delta T_i - \delta T_j)c_{ij}}{\sum_{ij}c_{ij}^2},
\end{equation}
where
\begin{equation}
    \delta T_i(\hat{n}_i,z_i) = T(\hat{n}_i) - \overline{T}(\hat{n}_i,z_i,\sigma_z).
\end{equation}
$T(\hat{n}_i)$ is the kSZ signal, that can be $T_{disk}$, $T_{AP}$, or $T_{MF}$.  The subtraction of $\overline{T}(\hat{n}_i,z_i,\sigma_z)$  given by:
\begin{equation}
    \overline{T}(\hat{n}_i,z_i,\sigma_z)=\frac{\sum_j T(\hat{n}_i)w(z_i,z_j,\sigma_z)}{\sum_jw(z_i,z_j,\sigma_z)},
\end{equation}
mitigates a potential redshift-dependent systematic that could contaminate the pairwise signal
where
\begin{equation}
    w(z_i,z_j,\sigma_z) = \exp\left(-\frac{(z_i-z_j)^2}{2\sigma_z^2}\right).
\end{equation}
with $\sigma_z$ = 0.01, following \citep{Li:2017uin}. 

We measure the pairwise signal in fifteen comoving pair separation bins of equal width between 0 and 150 Mpc, and four additional bin edges on 200, 250, 315, and 395 Mpc in accordance with \citep{Calafut:2021wkx}. The covariance of the pairwise signal is estimated with the bootstrap re-sampling analysis during which we randomly replace the peculiar velocities(or temperature decrements), v (or $\delta T$), of galaxy positions with replacement. We then estimate the covariance matrix, $C_{ij}$, by repeating this process 1000 times and then calculating the covariance of the list of pairwise momentum estimators calculated from each of the new samples.

\begin{figure*}
\includegraphics{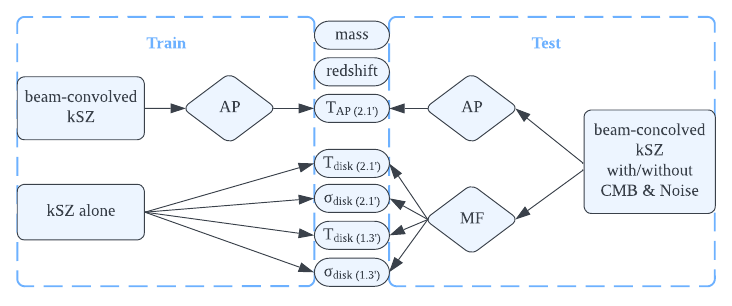}
\caption{The feature engineering process and the features used for our machine learning model for [left] the training and [right] the testing datasets. We consider seven features for each halo object, i.e. the mass and redshift, $z$, of each halo, the mean and standard deviation of disk kSZ temperatures, $T_{disk}$ and $\sigma_{disk}$, and the aperture photometry (AP) temperature, $T_{AP}$, within different angular scales (i.e. 1.3 and 2.1 arcminutes) from the halo center. For the test data, we use a matched filter (MF) approach to estimate the disk temperature properties, and consider scenarios with both pure kSZ and also data which includes primary CMB and detector noise.}
\label{fig:FE}
\end{figure*}

\section{Feature Engineering and Learning Model}
\label{sec:FE_LM}
In this section, we present the machine learning model used to recover the pairwise peculiar velocity. Machine learning, or statistical learning, is a field of study that utilizes a large amount of data for the computer to learn and understand the inherent patterns and interconnections in data and thereby make predictions and inferences without knowledge of the underlying physical laws or processes. The inputs that machine learning models utilize for training and inference to make predictions are known as model features. Model features are important as the accuracy of a machine learning model depends on the precise selection and composition of the features. The process of extraction and transformation of the features from raw data is called feature engineering. The motivation of feature engineering is to use these features to enhance the quality of the inference results as opposed to simply providing raw data to the machine learning algorithm.

One central goal of predictive modeling is to build a model that can make accurate predictions on new data that the model has never seen. In machine learning, data leakage refers to the situation when the model training process uses information that would not be expected to be available at the time of prediction. This happens when the model includes information on test data in the training process, and as a result, it achieves high performance while testing but might perform poorly for unseen data.

In this work, we use machine learning to predict the peculiar velocity of each individual cluster using features measured from the kSZ signal. 
We consider a sample of 300,000 cluster-mass halos randomly extracted from the Flender simulations spanning $0<z<1$ and $10^{13}M_{\odot}<M_{500}<3 \times 10^{15}M_{\odot}$, and associated kSZ maps, for training and testing purposes. To avoid data leakage, we use a train-test split procedure in which we train our model with 70\% of the sample (the ``Training Sample" of 210,000 halos) and use the remaining 30\% of them (``Flender Test Sample I" of 90,000 halos) \cite{Wang:2020kvd, Gholamy2018Why7O}, which the training model has never seen, for testing and model validation. In this way, we simulate how the model would perform on new, unseen data. 

For velocity reconstruction when detector noise and primordial CMB are present, we examine a second testing dataset from the Flender data (``Flender Test Sample II'') with a higher minimum mass threshold, as will be motivated in the next section. This sample consists of 86,866 Flender simulated halos $0<z<1$, $8 \times 10^{13}M_{\odot}<M_{500}<3 \times 10^{15}M_{\odot}$ drawn separately from Test Sample I from the Flender simulation. 

To facilitate the assessment of the approach beyond purely training and testing on a single simulation dataset,  we consider a dataset distinct from the Flender data, using the Websky simulations. Specifically, we use a randomly extracted sample of 300,000 cluster-mass halos spanning $0<z<1$ and $10^{13}M_{\odot}<M_{500}<3 \times 10^{15}M_{\odot}$ split 70\%/30\% into distinct training and testing subsamples.  We also consider a fourth separate test sample for analysis with primary CMB and noise of 90,000  Websky simulated halos (the ``Websky Test Sample"), with $0<z<1$ and $8 \times 10^{13}M_{\odot}<M_{500}<3 \times 10^{15}M_{\odot}$. 

These studies, and the use of $\sim$ 90,000 cluster test samples in our analysis, are motivated in the anticipation of BOSS and DESI data providing spectroscopic redshifts for photometrically selected galaxy clusters in the SDSS survey. Specifically, we use clusters of the Wen Han Liu (WHL) catalog\citep{Wen:2015uqa} as a guide. This catalog covers a redshift range of $0.05 < z < 0.75$ with a minimum mass of $5 \times 10^{13} M_{\odot}$ and consists of 158,103 galaxy clusters with photometric redshifts. Spectroscopic redshifts from BOSS are available for 121,103 (77$\%$) of the galaxy cluster sample, and of these 102,033 have $M_{500}$ $>8\times 10^{13}M_{\odot}$, with the expectation that DESI will provide additional spectroscopic redshifts. The full-sky Planck temperature maps would naturally have the same sky coverage as this catalog but the Planck map doesn't have enough resolution to do a matched filter\citep{Gong:2023hse}, a key process for our model that will be discussed below. On the other hand, the  ACT maps with higher resolution, do allow a matched filter approach, but they have a smaller sky coverage overlap with the WHL catalog. The motivation for developing the machine learning approaches here is to apply the approach to a spectroscopic catalog similar to WHL in combination with an upcoming multifrequency CMB data, such as from SO, CMB-S4, and CCAT, that is of higher resolution and higher sensitivity than Planck but will also survey a significant fraction of the sky.

For each cluster, we create seven features for training and testing purposes,  summarized in Fig.~\ref{fig:FE}. Given the peculiar velocity is directly proportional to the kSZ temperature, as in (\ref{eq:kSZ}), we consider the kSZ temperatures as our key features: $T_{Disk}$ in the training data (and $T_{MF}$ in the testing data) and $T_{AP}$. For $T_{disk}$ (training dataset) and $T_{MF}$ (test dataset) the mean and the standard deviation of the temperatures of the pixels within the aperture are used as input features.

These temperature features are directly measured from the observation map at the location of each cluster. For the $T_{disk}$ and $T_{MF}$ data we consider two different aperture sizes for the disk kSZ temperatures: where the signal-to-noise ratio (SNR) peaks at 2.1$^\prime$ \citep{Flender:2015btu, Gong:2023hse} and also a 1.3$^\prime$ aperture size that focuses more on the central peak kSZ signals. We also include a spectroscopic redshift and mass estimate for each halo in the set of features. In this work, we use the cluster mass estimate itself as the feature rather than a feature related to the mass proxy. Observationally, the cluster mass estimate could be inferred from one or more potential observable proxies of cluster mass, such as using richness \citep{Andreon:2016eck, Geach:2017crt}, X-ray \citep{Stanek:2006tu, Ettori:2013tka, Amodeo:2016wtq}, and weak lensing observations \citep{Applegate_2014, DES:2018kma, Murray:2022iyi}.  We discuss the potential effect of inaccuracies in the mass estimate on the modeling in more detail in section~\ref{sec:bias_mass}.

For training purposes, one can choose to train the model with either noise-free or noisy data.  Noise-free data enables the machine learning models to concentrate on the underlying structures and patterns we seek to extract. We have considered models trained on noise-free data and data including primary CMB and noise but find the latter perform significantly less well in reconstructing the features. Given this, and following the approach frequently employed with the machine learning technique, as listed in section \ref{sec:intro}, we train our model on the kSZ signal alone, without primary CMB and noise. Aperture photometry features are measured from the beam-convolved kSZ map, and the disk features are measured directly from the unfiltered pure kSZ map as summarized in the left of Fig.~\ref{fig:FE}. Note that we convolve the kSZ map with a beam for AP to account for the beam effect in real data analysis. 

We test our training model in two scenarios, one with a pure kSZ signal and the other when CMB anisotropies and detector noise are present. Similar to the training process, aperture photometry features are again measured from the kSZ map with beam convolution.  Different from the training process, the disk temperature features are measured using the matched filter from the kSZ map with beam convolution as summarized in the right of Fig. \ref{fig:FE}. The intent of using the matched filter and the beam-convolved kSZ map during the testing process is to mimic the signal extraction process when analyzing the real observation data. When adapting to real observations (e.g. when detector noise and CMB anisotropies are present), following \citep{Tanimura:2022fde}, the CMB-dominated signals above 30 arcminutes are filtered out, and the small-scale kSZ-related signals below 15 arcminutes are preserved, before we measure kSZ temperatures with AP and MF.

\begin{figure*}
\includegraphics[width = \textwidth]{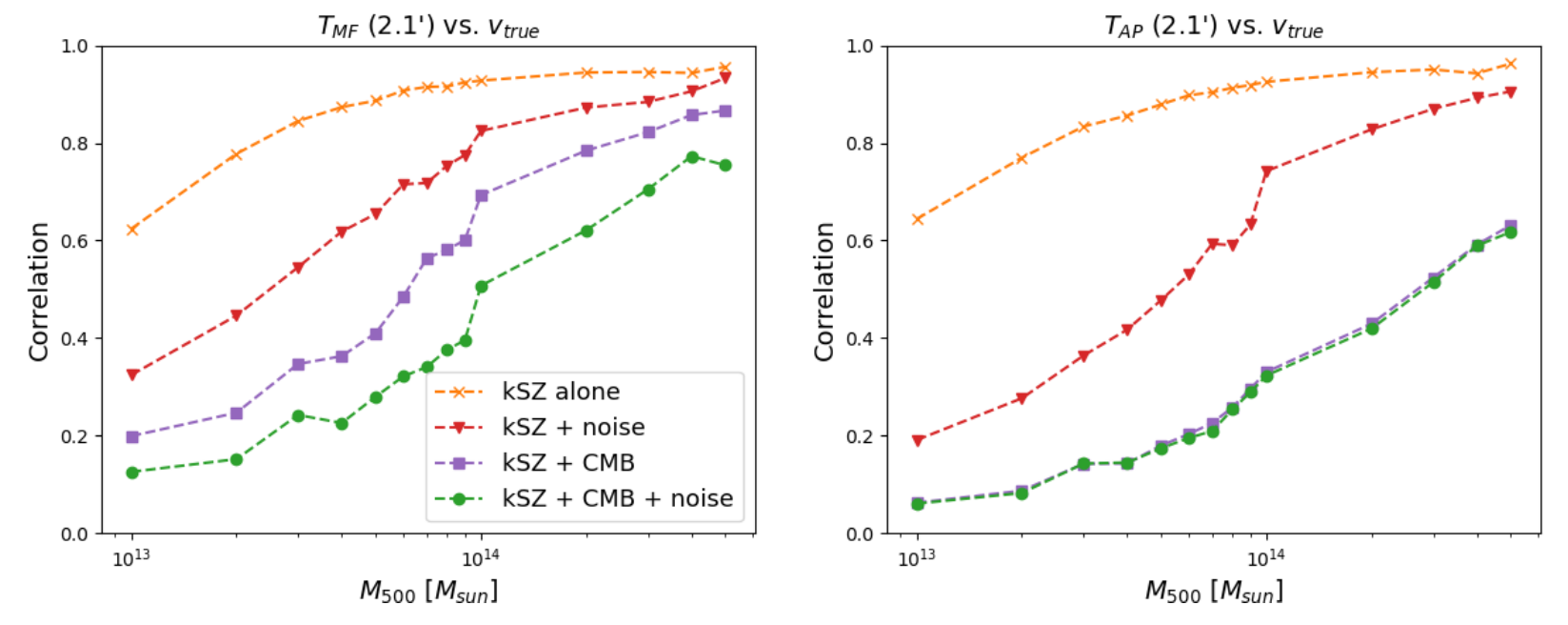}
\caption{The correlation between two of the key features the matched filter, $T_{MF}$ (2.1$^\prime$) [left] and aperture photometry $T_{AP}$ (2.1$^\prime$) [right], and the true peculiar velocity of each halo in Flender Test Sample I as a function of halo mass, when the test data includes just pure kSZ data [orange, cross], kSZ plus detector noise [red, triangle], kSZ plus primary CMB  [purple, square] and kSZ, primary CMB and detector noise [green, circle]. Note large-scale signals are filtered out before employing the MF and AP filters.
}
\label{fig:correlation}
\end{figure*}

This work aims to recover the pairwise peculiar velocity of the cluster sample through a one-step machine learning inference of estimating the individual peculiar velocity for each cluster in the testing sample. We train the model to predict a peculiar velocity for each cluster based on the seven features described above. This process is akin to a regression problem,  making an inference based on the numerical model features. We adopt a gradient boosting algorithm, LightGBM \citep{ke2017lightgbm}\footnote{\url{https://lightgbm.readthedocs.io/en/stable/}}, that has demonstrated stability and out-performance in making regression inferences with a set of 1-dimensional numerical model features. The gradient boosting model is in the form of an ensemble of weak learners, where a weak learner usually means a simple model that is slightly more accurate than random prediction. The model is trained sequentially, and in each iteration, a new weak learner is trained and added to the ensemble to correct for the previous model. In this process, several weak learners are combined into strong learners, in which each new weak learner is trained to minimize the prior model's loss, specifically the mean squared error is used here. 

One of the main challenges during the training process is to avoid over-fitting while achieving model accuracy and convergence. Below, we will briefly discuss our choice of two of the main hyper-parameters of the LightGBM model, i.e. the learning rate and the number of iterations. The learning rate determines how much the model corrects for the error at each iteration towards achieving a minimum of the loss. While a learning rate that is too low will take a long training time to converge and stuck in an unwanted local minimum, a learning rate that is too high will cause the learning to bounce over minima and never converge. The number of iterations, on the other hand, controls how many times the model corrects for the error. Too many iterations will usually cause an over-fitting of the training data that the model provides precise forecasts for training data, but not for unseen data, while too few iterations will not let the model learn sufficiently. To achieve higher accuracy, a common strategy is to let the model learn over more iterations and have a low learning rate. This is equivalent to finding a good combination of iterations and learning rates. In this work, we choose a learning rate value of 0.03 which is commonly used in many applications. We find that the model begins to converge after 3,000 iterations and starts to overfit after 5,000 iterations. We stopped our models at 3,600 iterations to avoid overfitting while also achieving high accuracy. The other hyperparameters of the model are determined using a grid search cross-validation technique running Scikit-Learn’s GridSearchCV \citep{scikit-learn}. Cross-validation is a technique that splits the dataset into subsets, or folds, with the purpose of using each fold as a validating set and the remaining folds for training. We evaluate the model performance on the validation set, where each fold is used exactly once as the validation set during the iterations of this process. The results from each iteration are then averaged to obtain the performance of the model. In the GridSearchCV algorithm, a grid of hyperparameters with possible values is first defined. Cross-validation is then used to train the model and assess its performance for each combination of hyperparameters as it explores the hyperparameter space. In this way, it helps in determining the ideal hyperparameter combination that yields the best model performance.

In assessing the efficacy of the machine learning velocity reconstruction, we utilize the following statistics.
We measure the strength of linear correlation between variables using the Pearson correlation coefficient. For a pair of random variables (X,Y), the correlation coefficient can be written as:
\begin{equation} \label{eq:correlation}
    \frac{\sum_{i = 1}^n (x_i - \Bar{x})(y_i - \Bar{y})}{\sqrt{\sum_{i = 1}^n (x_i - \Bar{x})^2} \sqrt{\sum_{i = 1}^n (y_i - \Bar{y})^2}},
\end{equation}
where $x_i,y_i$ are the individual sample data points, $n$ is the sample size, and $\Bar{x},\Bar{y}$ are the sample mean.

The signal to noise ratio (SNR) for the machine-learning model predicted pairwise velocity signal, is evaluated as
\begin{equation}
SNR = \sqrt{\sum_{ij}\hat{V}_{i,pred}C_{ij}^{-1}\hat{V}_{j,pred}}
\end{equation}
with $\hat{V}_{i,pred}$ the predicted pairwise velocity estimator using the machine learning model for the $i^{th}$ pair separation bin, and $C^{-1}_{ij}$ is the inverse of the covariance matrix estimated off a bootstrap of the dataset as discussed in \ref{sec:Vhat}.

We determine how well the pairwise velocity predicted by the  model fits to that obtained directly from the true velocities in the simulation with the $\chi^2$, 
\begin{equation} \label{eq:chi_v}
    \chi^2 = \sum_{ij} \Delta \hat{V}_i\,{C}_{ij}^{-1}\,\Delta\hat{V}_j,
\end{equation}
with $\Delta \hat{V}_i = \hat{V}_{i,true} - \hat{V}_{i,pred}$,
where $\hat{V}_{i,true}$ is the true pairwise velocity estimator for the test data. The best-fit model is achieved with the minimum $\chi^2$.

It is important to note how the $\chi^2$ should be interpreted in this analysis. Usually, we are comparing the fit of a theoretical model to observational survey data which can be viewed as a random realization of that model, modulated by cosmic variance and measurement uncertainties. In that case, a good fit to the data is characterized by a reduced $\chi^2$, $\chi^2$ per degree of freedom, value of around one as long as the covariance has been accurately estimated. In this analysis, however, we are comparing how well the machine learning model predicted velocity statistics for the test data compared with that true velocity correlation. In the limit of perfect reconstruction, the $\chi^2$ would be zero. The degree of variation between the predicted and true signals is affected by the intrinsic uncertainties in the {\it training} model data set and the efficacy of the training algorithm to recreate the test data set. The training data set used here is over twice the size of the test data and will have commensurately smaller statistical uncertainties than the test data. For the model to work well, the prediction uncertainties, $\Delta \hat{V}_i$, should be substantially smaller than the statistical uncertainties in the test data, characterized by the covariance $C_{ij}$. As such, the prediction from well performing machine learning algorithm should have a $\chi^2$ per degree of freedom that is well below one. 

\section{Analysis and Results}
\label{sec:results}

In section \ref{sec:vhalo} we consider the relation between the model features and peculiar velocities, and test the performance of the machine learning model to recover the individual halo velocities. In section \ref{sec:v_pairwise} we consider the pairwise velocity predicted using test data with solely the kSZ signal and, more akin to real observations, when detector noise and primordial CMB are included. We discuss the effects of systematic errors from cluster miscentering and mass misestimation in section \ref{sec:bias}. Finally, in section \ref{sec:cross sim} we consider the performance of the machine learning model when trained and tested using different kSZ simulations. 

\begin{figure}
\includegraphics[width = \columnwidth]{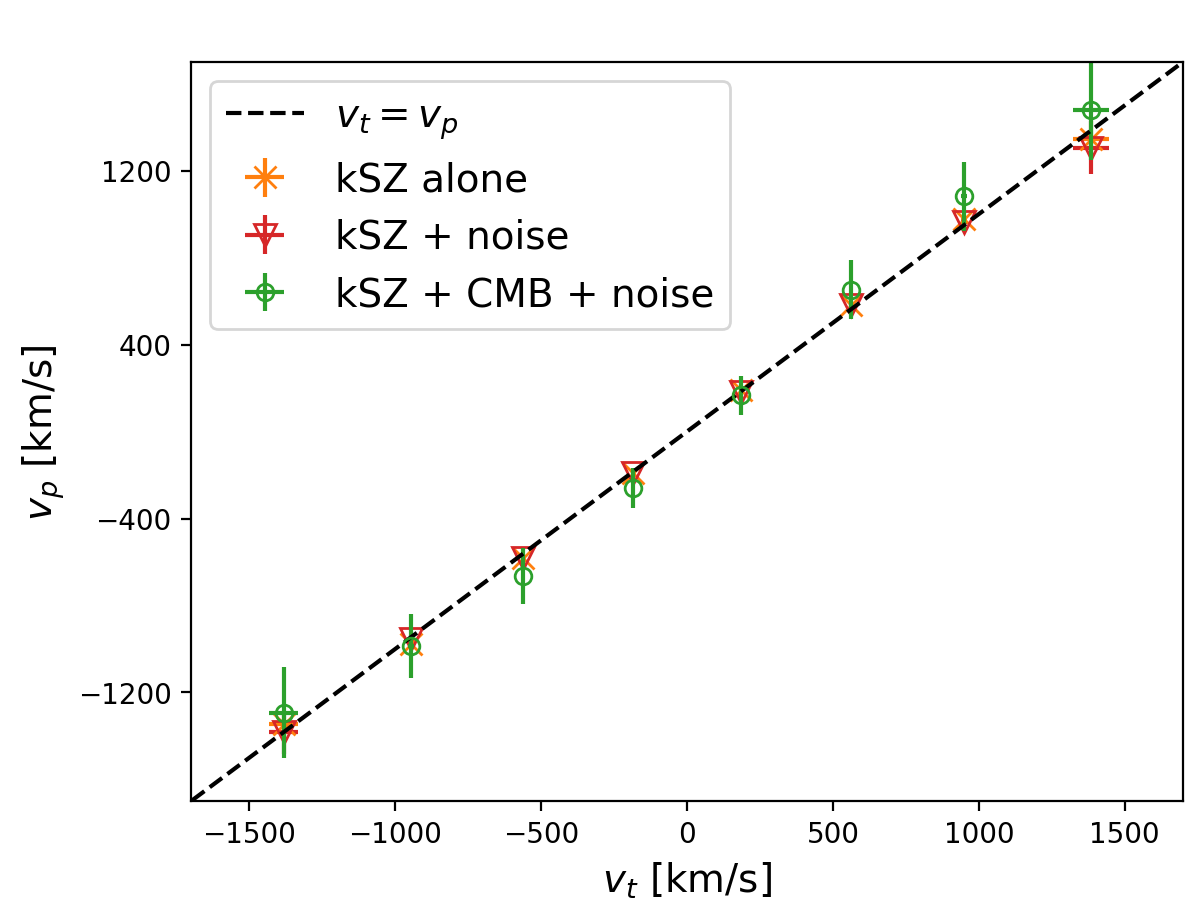}
\caption{A comparison of the machine learning predicted peculiar velocities, $v_p$, to the true peculiar velocities, $v_t$, for the Flender Test Sample II halo catalog, when the test data includes just kSZ [orange, cross], kSZ and detector noise [red, triangle], and kSZ and both primary CMB and detector noise [green, circle]. The mean signal binned by $v_t$ is shown with $1\sigma$ standard errors.} 
\label{fig:v_all}
\end{figure}

\begin{figure}
\includegraphics[width = \columnwidth]{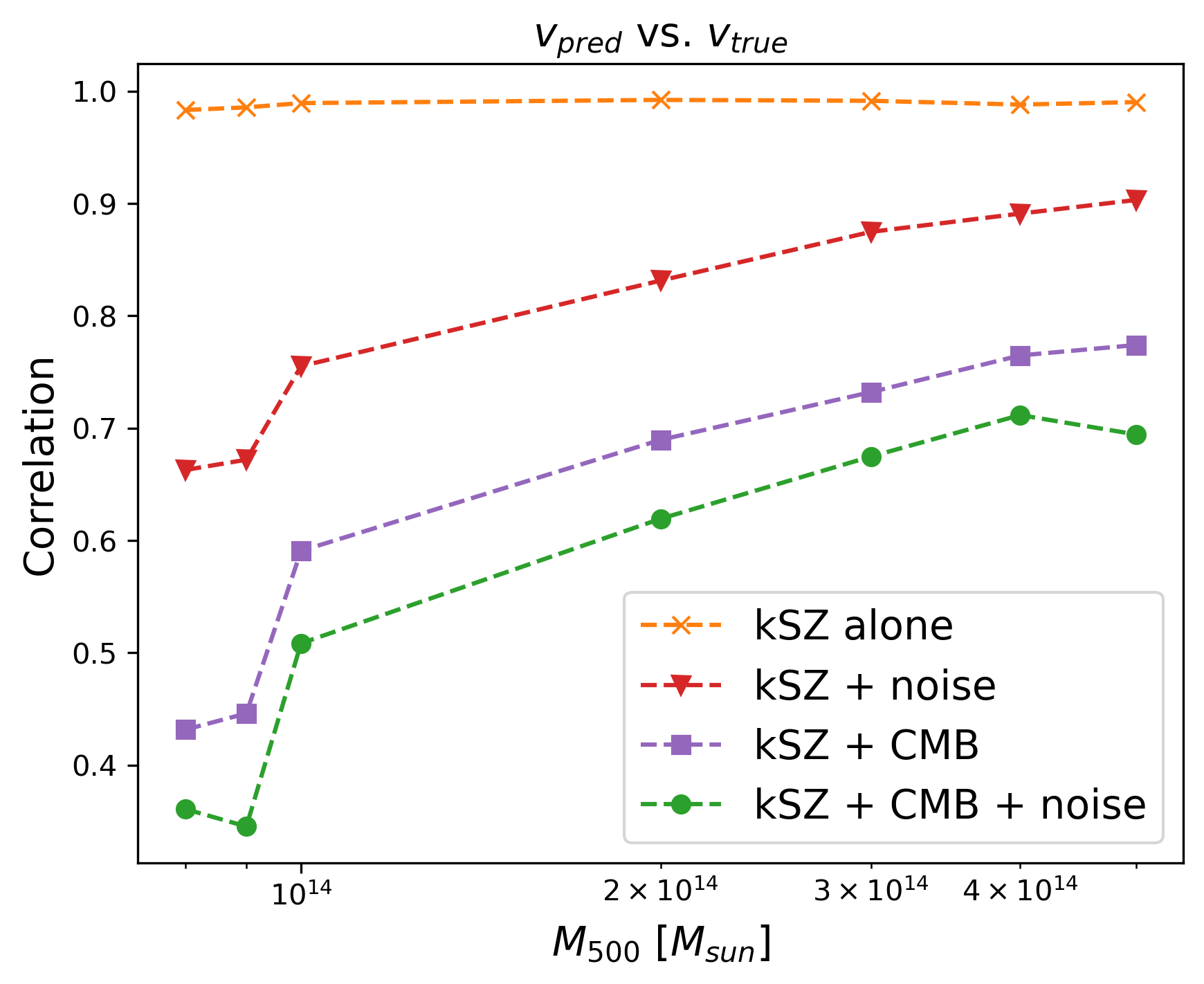}
\caption{The correlation between the predicted and true velocities, $|v_p|$ and $|v_t|$, as a function of halo mass for Flender Test Sample II, when the test data includes only kSZ [orange, cross], kSZ+detector noise [red, triangle], kSZ+primary CMB [purple, square] and kSZ, primary CMB and detector noise[green, circle].}
\label{fig:vcorrelation}
\end{figure}

\subsection{Individual halo velocity reconstruction}
\label{sec:vhalo}

In advance of developing the machine learning model, we first establish how the key model features relate to the halo's peculiar velocity. In Fig.~\ref{fig:correlation}, we show the correlation coefficient between two of the key features, $T_{MF}$ (2.1$^\prime$) and $T_{AP}$ (2.1$^\prime$), and the true peculiar velocity of each halo, $v_t$,  as a function of the halo mass for four kSZ datasets that include solely the kSZ, and combinations of the primary CMB and detection noise. 
 
When just the kSZ is included the correlation of the MF and AP temperatures with the true velocity are both high and consistent with one another. For the highest mass halos, the correlation is nearly perfect. For lower mass halos the correlation falls, to just above 0.6 for masses of $10^{13}M_{\odot}$. This is consistent with the kSZ signal being smaller for lower mass clusters and thus the relative contribution from other kSZ sources along the line of sight, but unrelated to the cluster mass halo, becomes more important, creating the noisier signal.

When detector noise and/or primordial CMB are included in the test data, there is, as is to be expected, a weaker correlation between the measured model features and the true peculiar velocity. We find that, in general, kSZ signals extracted with matched filter have a tighter correlation with the cluster peculiar velocity than those extracted with aperture photometry. The inclusion of the primary CMB has the largest impact on weakening the correlation for both signal extraction methods. 

Using a correlation of around 0.5 as a guiding threshold  \citep{Ratner2009TheCC}, we find that an overall correlation of 0.46 between $T_{MF}$ and the true peculiar velocities is maintained for a test sample with minimum mass threshold $M_{500}=8 \times 10^{13} M_{\odot}$. By contrast, a sample needs to have $M_{500}> 3 \times 10^{14} M_{\odot}$ for $T_{AP}$ to retain a correlation of 0.5 of higher. We seek to include as many halos as possible to minimize statistical uncertainties and maximize performance. Given the poorer correlation of the AP features, and the restrictive mass threshold this would impose when the detector noise and primordial CMB are present, we exclude the AP temperature features (i.e. $T_{AP}$) and preserve only the MF-filtered kSZ temperature features(i.e. $T_{MF}$).

Motivated by this, we create a second halo testing sample, Flender Test Sample II  with $M_{500} > 8 \times 10^{13} M_{\odot}$, containing 86,866 objects distinct from those used in the training model. Fig.~\ref{fig:v_all} shows the comparison of the predicted velocity using the machine learning model with only the matched-filtered kSZ temperature inputs from the test sample to the true velocity for this sample. We find that the application of the machine learning model allows us to recover the peculiar velocities without bias for both kSZ-only signals and when the detector noise and primordial CMB are added.

In Fig.~\ref{fig:vcorrelation} we show the correlation between the halo velocity predicted by the machine learning model, $|v_p|$, and the true velocity, $|v_t|$, as a function of mass for Flender Test Sample II with four different kSZ datasets: kSZ alone, and when noise and primary CMB are included. We find that the trends in correlation are consistent with those observed between kSZ temperatures and the true velocity shown in Fig.~\ref{fig:correlation}. The correlation for the overall correlation across all mass bins is 0.43, reinforcing that $8\times 10^{13} M_{\odot}$ is a reasonable minimum mass threshold for the analysis.  

In the rest of the paper, we will focus the analysis on the samples with $M_{500} > 8 \times 10^{13} M_{\odot}$ and also exclude the AP temperature features as discussed above. 

\subsection{Pairwise peculiar velocity reconstruction}
\label{sec:v_pairwise}
\begin{figure}
\includegraphics[width = \columnwidth]{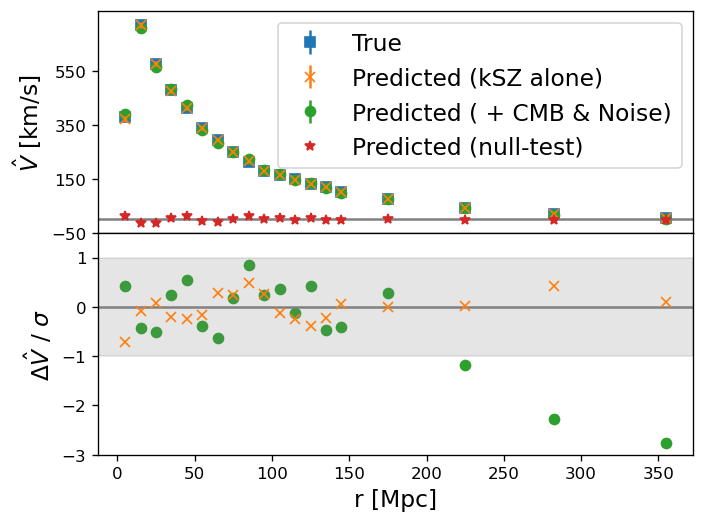}
\caption{[Upper] The pairwise peculiar velocity estimator calculated for the true velocity ($\hat{V}_t$) [blue, square] and predicted velocity ($\hat{V}_p$) with pure kSZ signals [orange, cross], the mean of 10 realizations when detector noise and primary CMB are included [green, circle], and the null test of using just the CMB and detector noise (and no kSZ) [red, star] for Flender Test Sample II. [Lower] The difference between prediction and expectation (e.g. $\Delta \hat{V} = \hat{V}_p - \hat{V}_t$) scaled relative to the respective uncertainty, $\sigma$, for each sample.}
\label{fig:pairwise_pure_CMB}
\end{figure}

Using Flender Test Sample II, we start by testing how well we can recover the pairwise peculiar velocity using the pure kSZ signals in the test set model features. Fig.~\ref{fig:pairwise_pure_CMB} shows the pairwise velocity statistics associated with the predicted individual halo velocities described in \ref{sec:vhalo}. We find that the predicted statistic, $\hat{V}_p$, recovers the true estimator, $\hat{V}_t$,  well within the statistical errors estimated from bootstrapping the sample. Comparing the reconstructed pairwise signal to the true one gives a best-fit $\chi^2 = 1.6$ for 19 bins covering all the pair separations to 395 Mpc.

We then test the feasibility of our machine learning model to extract the estimator from more realistic observations, when both primary CMB and detector noise are included which all together can be orders of magnitude larger than the kSZ signal alone. 

We generate 10 realizations of both the detector noise and primary CMB as described in section \ref{sec:data}. We generate independent primary CMB and detector noise realizations from each of their individual power spectrum without kSZ included. Thus, there will be no correlation between these ten
realizations. For each of these 10 realizations, primary CMB is added to the Flender model 3 map before a beam convolution of FWHM = 1.4$^\prime$, after which the detector noise is included. As mentioned in section \ref{sec:FE_LM}, the large-scale CMB-dominated signals are filtered out before the model features are created for each halo object. The mean pairwise peculiar velocity estimator across the 10 realizations is calculated and errors are estimated by conducting a bootstrap on one realization. 

The results are shown in Fig.~\ref{fig:pairwise_pure_CMB}. We first conduct a null test in which the mean signal is obtained for purely the 10 maps of detector noise and primary CMB with no kSZ. The reconstructed signal is consistent with $\hat{V}=0$.  When the kSZ signal is included, the machine learning model recovers an unbiased estimate of the pairwise peculiar velocity for cluster pair separations below 200 Mpc. For the 16 bins with $r<200Mpc$, we find a best-fit $\chi^2 = 9$ for the mean pairwise signal over the 10 realizations and SNR = 11.  For separations greater than 200 Mpc, the intrinsic amplitude of the pairwise signal is very small. While the signal is able to be recovered for the pure kSZ inputs, once detector noise and primary CMB anisotropies are included, and dwarf the kSZ signal in the test features, the machine learning algorithm is unable to recover a robust prediction of these low amplitude signals; the inferences are systematically biased for these large pair separations. For this reason, the analyses in the rest of the paper focus on the pairwise separations below 200 Mpc.

The machine learning inference presents an alternative to the approach in which the pairwise velocity estimator is reconstructed by combining the pairwise kSZ momentum with associated optical depth estimates. The pairwise kSZ momentum estimator calculated from the MF temperature features has SNR = 11. If we consider an optical depth estimate using tSZ or X-ray observations with an assumed $18\%$ uncertainty \citep{Vavagiakis:2021ilq, Flender:2016cjy}, we would estimate a pairwise velocity reconstruction with SNR = 7, compared to SNR=11 using the machine learning method in this work.

\begin{figure}
\includegraphics[width = \columnwidth]{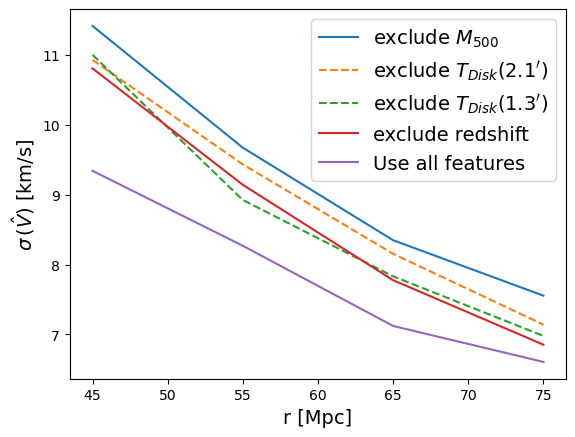}
\caption{Sensitivity of the $1\sigma$ standard errors on the pairwise peculiar velocity momentum derived from the machine learning algorithm when trained and tested without one of the features included. The cluster feature set that excludes $M_{500}$ [blue], $T_{Disk}(2.1')$ [orange], $T_{Disk}(1.3')$ [green], redshift [red], and no feature exclusion [purple] are shown for Flender Test Sample II.}
\label{fig:feat_importance}
\end{figure}

To understand how the model learns to break the degeneracy between velocity and optical depth from the kSZ temperature, we have tested the feature importance by excluding one feature at a time and evaluating the model performance after the exclusion of each feature. Note that each time we re-train and test the model using solely the selected features. In Fig. \ref{fig:feat_importance}, we show the $1\sigma$ standard errors for $40<r<80$ Mpc for the pairwise velocity statistics derived from the predicted peculiar velocity when each feature is excluded from the model. The exclusion of the halo mass in the training data gives the largest increase in the 1$\sigma$ errors on pairwise velocity statistics. This suggests that the mass information plays an important role in velocity estimation model consistent with the expectation that the mass feature serves as a proxy for the optical depth estimate and can therefore improve the model accuracy through an implicit cluster $M_{500}-\tau$ relation.

\subsection{Effects of uncertainties in the test data}
\label{sec:bias}
In this section, we discuss how the machine learning model performs when accounting for potential systematic errors arising due to cluster miscentering and mass misestimation. 

\subsubsection{Miscentering bias} 
\label{sec:bias_centering}
While in simulations we can identify the gravitational center of a cluster halo, observationally this is not the case. Typically, following the `Central Galaxy Paradigm' \cite{vandenBosch:2005tk}, the most massive, and brightest, galaxy in the cluster will be the most centrally located. As such, the brightest galaxy identified in a cluster will commonly be used as the proxy to define the center of a cluster. The position of the brightest galaxy may not always be the gravitational potential minimum of the cluster, however, and the brightest galaxy may also not be the closest cluster member to the center. Roughly 30\% of cluster mass systems, in halos $10^{13}-10^{14.5}M_{\odot}$, are found to not have the brightest galaxy as the central most member  \cite{Skibba:2010ez,Rozo:2014sla,Hoshino:2015zza,DES:2015mqu,Lange:2017dna}.  Hence, the cluster center estimated from the brightest galaxy can be biased from the true cluster center. This can lead to biases in the calculated pairwise signal kSZ \cite{Calafut:2017mzp} as well as other cluster measurements such as in the X-ray \cite{Seppi:2022mdo}.

Saro et al. \citep{DES:2015mqu} compared the cluster center derived from the tSZ profile to the position of the brightest galaxy and found that 63\% of clusters have a miscentering of $\sigma_0 = 0.07 h^{-1}$ Mpc, and the rest have a miscentering of $\sigma_1 = 0.25 h^{-1}$ Mpc. This miscentering can be expressed in terms of a bimodal Gaussian distribution,
\begin{equation} \label{eq:bimodal}
P(x) = 2\pi x \, \left(\frac{\rho_0}{2\pi \sigma^2_0}\exp\left\{-\frac{x^2}{2\sigma^2_0}\right\} + \frac{1-\rho_0}{2\pi \sigma^2_1}\exp\left\{-\frac{x^2}{2\sigma^2_1}\right\}\right),
\end{equation}
with $\rho_0 = 0.63$. 

\begin{figure}[!t]
\includegraphics[width = \columnwidth]{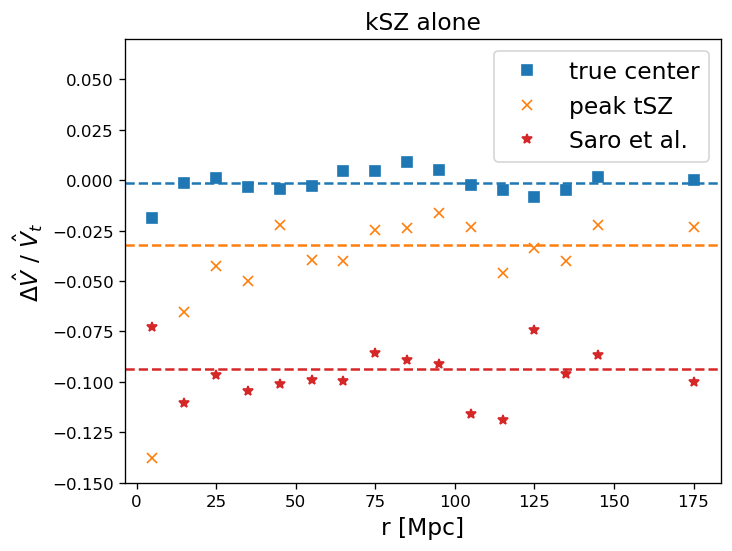}
\caption{The fractional difference ($\Delta \hat{V}/\hat{V}_t$) between the estimator ($\hat{V}_t$) derived from the true velocity and the prediction ($\hat{V}_p$) derived from the machine learning model for Flender Test Sample II when the test data is kSZ-alone. Predictions when the test model features are calculated assuming cluster centers modeled from the true center [blue, square], from displacements based on Saro et al. [red, star], and when the peak tSZ signal is used as the cluster center [orange, cross]. The mean signal deviation across the whole sample is also shown for each case [dashed line].}
\label{fig:miscenter}
\end{figure}

Following Saro et al., we create a modified version of Flender Test Sample II in which the test model features are created using halo centers which have a random offset applied using (\ref{eq:bimodal}). In Fig.~\ref{fig:miscenter}, we present the resulting predicted pairwise velocity statistics when test features are just for the kSZ alone (with no detector noise and CMB included). Miscentering at the level of the Saro model reduces the predicted pairwise peculiar velocity ($\hat{V}$) by 9\% relative to that derived using the true centers to derive the test data model features.

To alleviate the suppression of the amplitude of the pairwise velocity estimator caused by miscentering, we consider an alternative centering approach in which the center is located using the tSZ signal profile. The process is summarized as follows. We first assign each cluster a new center following the Saro et al. model. We then use the tSZ map and search around the new center to locate the peak tSZ signal. By following these steps we reflect the practical process that we would do with real data in which the initial guess of the center is determined by the Saro et al. model. We finally assign a center for each cluster based on the location of the identified peak tSZ signal and calculate model features using that assumed center. 

Using the tSZ-derived center, the predicted pairwise peculiar velocity ($\hat{V}$) is reduced by only 3\% on average; the associated results are shown in Fig.~\ref{fig:miscenter}. We conclude that using the cluster center derived from the peak tSZ signal could alleviate the suppression of the amplitude of the pairwise velocity estimator caused by miscentering if the optically brightest galaxies are purely used as the cluster center proxy. 

\begin{figure}[!t]
\includegraphics[width = \columnwidth]{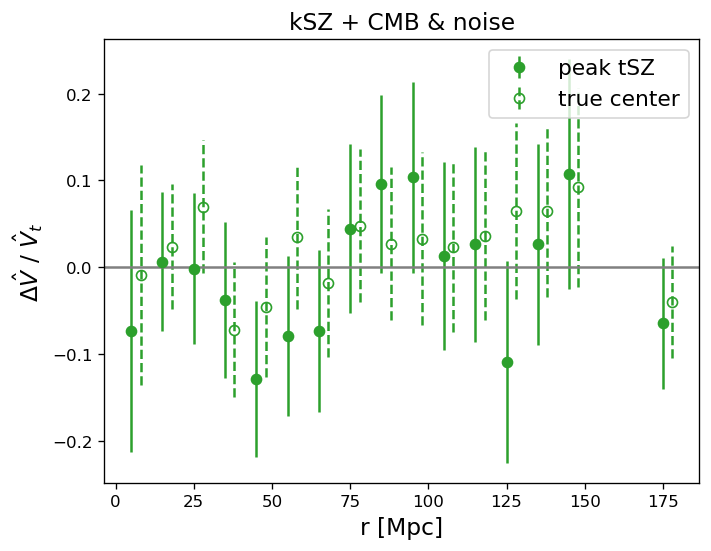}
\caption{The fractional difference ($\Delta \hat{V}/\hat{V}_t$) between the pairwise velocity estimator from the true velocity  ($\hat{V}_t$) and that from the machine learning prediction ($\hat{V}_p$) derived from kSZ and a single realization of the primary CMB and detector noise for Flender Test Sample II when the location of the peak tSZ signal is used as the proxy for the cluster center [filled circle]  and the true cluster center from the sims is used [empty circle]. 1$\sigma$ error bars estimated from a bootstrap of the test data are shown.}
\label{fig:miscentercmb}
\end{figure}

In Fig.~\ref{fig:miscentercmb}, using one realization of kSZ, CMB, and noise, we compare the machine learning pairwise $\hat{V}$ prediction derived from the peak tSZ center model to the true $\hat{V}$ derived from the true halo velocity provided in the catalog. We find that the inclusion of detector noise and CMB anisotropies does not additionally suppress the amplitude from the miscentering and that the offset due to miscentering is smaller than the sample's statistical errors. A best-fit $\chi^2$ of 11 is found for the predicted pairwise signal from the single CMB and noise realization, relative to the underlying true pairwise velocities when the centers are derived from the tSZ peak. This just is slightly larger than the best-fit $\chi^2=10$ when the true cluster centers are used for the test feature for the same noise and CMB realization. 

\subsubsection{Mass scatter bias}
\label{sec:bias_mass}
In addition to uncertainties related to the location of the cluster center, the cluster mass is also a model feature that will need to be estimated from observational proxies.
For a galaxy survey, a common approach, as used in \citep{Vavagiakis:2021ilq, Calafut:2021wkx}, is to estimate the stellar mass from the luminosity of the brightest galaxy using the mass-to-light ratio \citep{Chabrier:2003ki, Kravtsov:2014sra, Bell:2000jt, Bell:2003cj}, and then the halo mass is derived from the stellar-to-virial mass relation, $M_{\star} - M_{vir}$ \citep{Kravtsov:2014sra}. Other approaches to mass estimation include using the mass-richness relation, and calibration from combinations of X-ray,  weak lensing, and tSZ observations. Any uncertainty in these relationships could lead to added dispersion or bias between the halo mass estimated as the input feature and the true halo mass.

\begin{figure}
\includegraphics[width = \columnwidth]{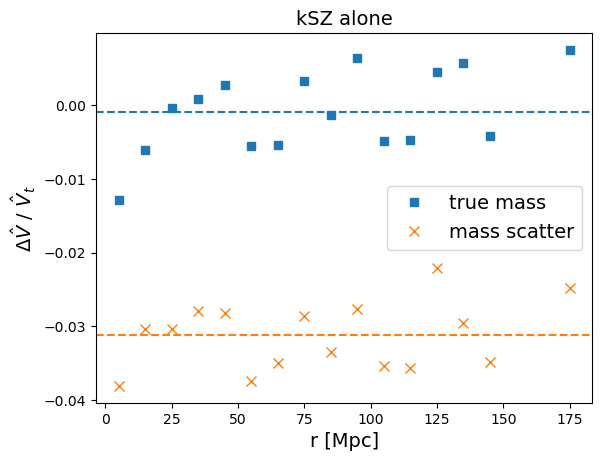}
\caption{The fractional difference ($\Delta \hat{V}/\hat{V}_t$) between the true estimator ($\hat{V}_t$) derived from the true velocity and the prediction ($\hat{V}_p$) derived from the machine learning model for pure kSZ test data for the augmented Flender Test Sample II. The predicted velocity when the true cluster masses are used as input features [blue, square] is shown along with that when the mass features incorporate scatter to reflect misestimation [orange, cross]. The mean fractional difference [dashed line] is also given for each sample.}
\label{fig:massscatter}
\end{figure}

We consider a scatter in the logarithmic mass estimate of 20\% for each halo mass, reflective of mass uncertainty estimates using mass-richness and weak lensing calibration approaches (e.g.\cite{Gruen_2011, Kohlinger:2015jna, SPT:2016gov, Baxter:2016jdq}). We use the Flender Test Sample II and augment it with a random subsample of lower mass halos down to $10^{13} M_{sun}$. Using the sample size of Flender Test Sample II as a reference, the attached subsample is selected with number distribution following the halo-mass function of the Flender simulation. In this way, we avoid attaching too few or too many lower-mass halos, and the combined subsample can have the same halo mass distribution as the whole catalog. For each cluster, we assign a new mass based on relation $ln \, M_{new} = ln \, M_{500} + \Delta$, where $\Delta$ is drawn from a Gaussian distribution with zero mean and standard deviation $\sigma = 0.2$. 

We first note that introducing this uncertainty has implications for the minimum mass threshold imposed to create the sample. The scattering results in some clusters in the original sample with true masses $M_{500}> 8 \times 10^{13} M_{\odot}$ being excluded based on the mass estimate; we find $87\%$ of the original sample objects are retained in the new sample. The new ``augmented Flender Test Sample II" will also include some clusters with true masses $M_{500}< 8 \times 10^{13} M_{\odot}$ for which the inferred mass now exceeds the threshold. Because the cluster mass function decreases as one goes to larger masses, we find that the overall test sample size, with the scatter included, would increase by 8\% as more lower-mass halos will move above the threshold than higher-mass ones will fall below. Given our model takes the cluster mass as an input feature, we start by testing how the mass scatter would affect the model prediction with pure kSZ signals when each cluster is assigned the `wrong' mass and this is used as an input feature. 

\begin{figure}
\includegraphics[width = \columnwidth]{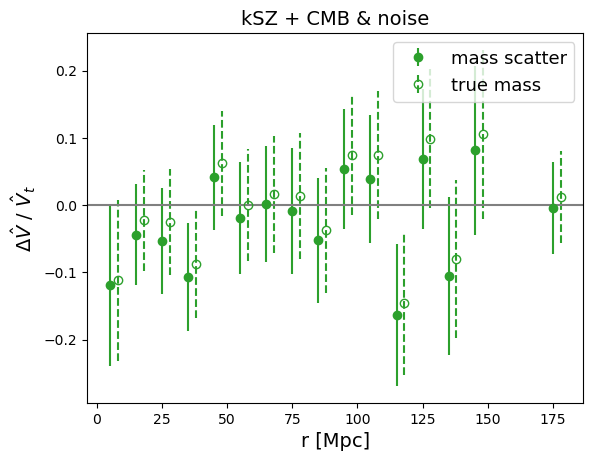}
\caption{The fractional difference ($\Delta \hat{V}/\hat{V}_t$)  between the pairwise velocity estimator from the true velocity ($\hat{V}_t$) and that predicted by the machine learning model ($\hat{V}_p$) derived from kSZ + a single realization of the primary CMB and detector noise for the augmented Flender Test Sample II. The velocity when the true mass is used as a model feature [empty circle] is compared to that when the mass feature is modified to account for mass misestimation [filled circle]. 1-$\sigma$ error bars estimated using a bootstrap of the test data are also shown. }
\label{fig:massscattercmb}
\end{figure}

In Fig.~\ref{fig:massscatter}, we compare the predicted $\hat{V}$ with the true $\hat{V}$ for the new augmented Flender Test Sample II. We find that mass scatter reduces the predicted pairwise signal by 3\% on average for the new sample.  This is understandable given the sample with the mass scatter will include lower mass halos, that had previously fallen below the mass threshold, for which the mass is overestimated. As such, with the higher `wrong' input mass as an input feature along with the kSZ temperature, with an amplitude consistent with its true lower mass, the machine learning model will predict a smaller absolute velocity on average. 

In section~\ref{sec:vhalo}, we demonstrated that the correlation between velocity and the input features falls below 0.5 for cluster samples with $M_{500} < 8 \times 10^{13} M_{\odot}$ when primary CMB and detector noise are included. Given the halos in our sample selected based on mass estimates incorporating scatter have true mass $M_{500} < 8 \times 10^{13} M_{\odot}$, we want to consider the effect of mass scatter when these lower mass halos and primary CMB and detector noise are present in the input features. In Fig.~\ref{fig:massscattercmb}, we find that the mass scatter does produce a systematic $\sim3\%$ reduction in the estimated pairwise velocity when the CMB and noise are included, relative to that inferred if the true halo masses are used as input features. The deviations are largely captured within the $1\sigma$ bootstrap uncertainties. 

The best-fit $\chi^2$ for the predicted pairwise signal from the single CMB and noise realization, relative to the underlying true pairwise velocities, is 11 for the 16 bins when the mass scatter is included, versus 10 when we assume perfect knowledge of the cluster masses.

Note here, to demonstrate the machine learning prediction efficacy, the comparisons above are between the predicted and true velocity correlations for the augmented Flender Test Sample II sample. Separately from this, the true pairwise velocity statistic for the augmented sample is in itself intrinsically 2.4\% lower, on average, than that for the original Flender Test Sample II, because the augmented sample has a lower mean mass. Both effects, the difference in the two samples and the bias in the machine learning prediction, each arising from the mass misestimation, would have to be factored in if the derived pairwise velocity is to be compared to theoretical cosmological predictions that assume a given cluster mass distribution.

\subsection{Application to other simulations}
\label{sec:cross sim}
To broaden beyond the Flender simulation, we also test the machine learning model on a second kSZ simulation, the Websky simulation \citep{Stein:2020its}. We upgrade the Websky kSZ map to the same resolution as the Flender simulation before we measure the model features, and we use the same signal template profile (parameters), calibrated based on FL3, for the matched filter across different simulations. 

Similar to the Flender simulation, we find that training on 70\% of the pure kSZ signals on the Websky simulation and testing it on the remaining 30\% of the Websky simulation gives an unbiased reconstruction of the pairwise peculiar velocity estimator. Note that we find the Websky trained machine learning model performs poorly for scales above 200 Mpc, because the systematic errors dominate, as it did for the Flender trained one. 

\begin{figure}
\includegraphics[width = \columnwidth]{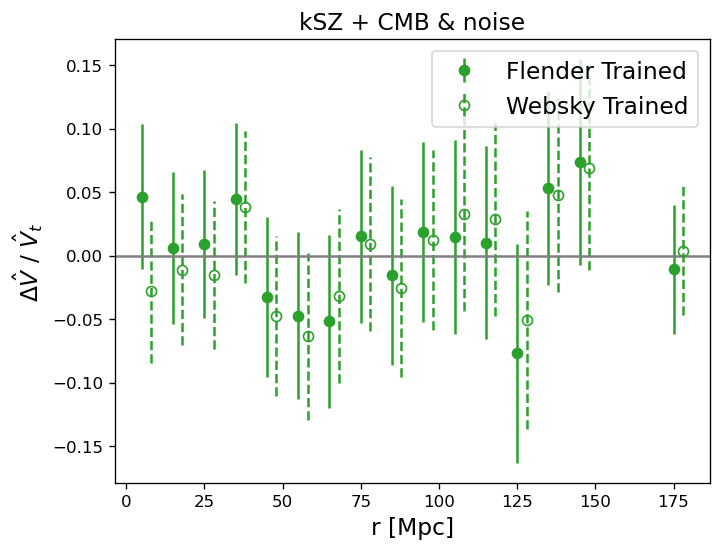}
\caption{The average fractional difference ($\Delta \hat{V}/\hat{V}_t$)  between the mean pairwise velocity estimator from the true velocity ($\hat{V}_t$) and that from the machine learning prediction ($\hat{V}_p$) for the Websky Test Sample over 10 realizations of the detector noise and primary CMB, using the machine-learning model trained on data from the Flender simulation [filled circle/full line] and trained on the Websky simulation [open circle/dashed line].}
\label{fig:Flender_on_Websky}
\end{figure}

We then cross-validate our model by applying the model trained on the Flender simulation to predict the velocities in the Websky Test Sample with a mass sample $M_{500}> 8 \times 10^{13} M_{\odot}$, as outlined in section \ref{sec:data}. The 10 realizations of the detector noise and primary CMB used on the Flender analysis were also included and a bootstrap analysis of one realization was performed to estimate the signal covariance.  We find that the machine learning algorithms respectively trained on the Flender simulations and Websky simulations can both recover the true $\hat{V}$ well for the Websky training sample as shown in Fig.~\ref{fig:Flender_on_Websky}. 

When the model is trained on the Flender simulations, the mean predicted pairwise signal over the 10 realizations for the Websky test data has a best-fit $\chi^2$ of 11 and an SNR of 9, for the 16 bins for $r<200$ Mpc. When trained on the Websky simulations, using the same 10 realizations of CMB and noise, the predicted signal has a best-fit $\chi^2$ of 11 and SNR of 10. The two predictions, trained on the different simulation sets, are therefore very consistent, showing that the approach is not limited to simply reconstruction of test data derived from the same simulations as the training set. This demonstrates the potential for using this machine learning model, trained on the Flender simulations (or other future high fidelity kSZ and galaxy simulations), as a suitable approach for the analysis of real survey data.

\section{Conclusion}
\label{sec:conc}
In this work, we have demonstrated that a machine-learning model using the gradient boosting algorithm trained on high-fidelity simulations can be used to recover an unbiased estimate of pairwise peculiar velocity from kSZ and halo catalog data. This provides a tantalizing alternative to the two-step process currently used to extract the velocity estimate, combining a pairwise kSZ momentum estimator with a separate estimate of the cluster optical depths, with the additional uncertainties that involves. 

For the machine learning approach, we consider seven halo features for training and prediction. Following \citep{Gong:2023hse}, we consider two signal extraction techniques,  matched filter and aperture photometry, to create the kSZ temperature features. We measure kSZ signals after filtering out CMB-dominated signals ($\ell < $  360) when adapting to real observations. 

We first test the feasibility of our machine learning model to recover the halo individual peculiar velocity. We find that the model is efficient at recovering an unbiased measurement of the individual peculiar velocity for both pure kSZ signals and when detector noise and primary CMB are added.  Given that the primary CMB dominates at large scales, and the detector noise dominates at small scales, the tSZ anisotropies and foreground contamions are sub-dominant than the primary CMB and detector noise. We thus did not test the presence of these noises in this work. We suggest testing the deprojection and the presence of these foreground contaminations in future work. While the presence of the CMB and noise makes the correlation weaker between temperature features and peculiar velocities we find that a minimum mass threshold of $M_{500} > 8 \times 10^{13} M_{\odot}$ is practical for confidently recovering the halo velocities with the model using a matched filter. The correlation is lower when using aperture photometry, which translates into a far higher minimum mass threshold. For this reason, aperture photometry features are not incorporated in the main analysis. At large separations, of $r>200Mpc$, the kSZ signal becomes very small, and non-kSZ signals dominate leading to inaccurate and biased predictions. For separations $r<200Mpc$, the machine learning approach provides an unbiased prediction of the velocities with modeling uncertainty levels that are significantly smaller than the statistical uncertainties estimated to be inherent in the test sample itself. We also test the model feature importance by excluding different features from the model and evaluating the model performance. We find that the mass information is the most important feature that break the degeneracy between velocity and optical depth from the kSZ signals. The mass serves as a proxy for optical depth through
an implicit cluster mass - $\tau$ relation that allows the model to differentiate among the different levels of kSZ temperature and velocity combinations.

We consider the effects of two potentially significant systematic effects: cluster location miscentering and mass misestimation. We find that using the location of the peak tSZ signal as the proxy to define the cluster center could reduce the systematic error in the positional offset introduced by using the brightest galaxy as a proxy for the halo center. The pairwise velocity amplitude suppression from miscentering is reduced from 9\%, when modeling centering using a brightest galaxy, to 3\% when using the peak tSZ signal as the proxy center. In considering the impact of cluster mass misestimation, we find that a 20\% mass scatter would reduce the predicted pairwise velocity estimator by 3\% for both pure kSZ signals and when detector noise and primary CMB are added. Such 3\% offsets fall well within the 1$\sigma$ statistical sample uncertainty estimates. 

While the main analysis uses the Flender simulations to train and test the machine-learning model, we also consider its performance with a second kSZ simulation to demonstrate its robustness and ability to be generalized. We use the Websky simulation which includes similar astrophysical effects to the Flender simulation while at a lower resolution. We find that the model trained on the Websky simulation is efficient at reconstructing the pairwise velocity estimator when applied to the Websky test data. Furthermore, the velocity statistics in the Websky test data are also able to be accurately reconstructed with the Flender-trained model.

In this work, we show the feasibility of using a machine learning model that uses cluster properties measurable with survey data as features to recover the pairwise velocity estimator with the intent to demonstrate its applicability to upcoming real observations.  We focus on one approach using a gradient-boosting algorithm, LightGBM. Given other machine learning applications to the analysis of galaxy clusters, for example, Neural Networks for the study of mass estimation\citep{deAndres:2022mox}, further opportunities for improved performance could well be realized by using additional model features together with a more complex feature engineering process. This could include enhancing the kSZ temperature features by including additional velocity information using velocity reconstruction with the galaxy number density field \citep{Guachalla:2023lbx, Hadzhiyska:2023nig} or additional cluster features, such as the tSZ, weak lensing, X-ray and richness statistics related to target clusters, to augment the cluster fitting model. Improvement of the mass calibration might also be achieved through machine learning approaches trained on large scale simulations to refine calibration/scaling relations. Novel multi-frequency techniques \citep{Madhavacheril:2019nfz, Han:2021vtm, McCarthy:2023hpa, Pratt:2024skt} have been developed to deproject the Compton-y and foreground contaminations, such as the cosmic infrared background(CIB), from the CMB observations. Such techniques have been used, for example, to provide a cleaned kSZ + CMB-only maps as the data products of ACT DR4 \citep{Madhavacheril:2019nfz}, analyzed in \citep{Vavagiakis:2021ilq, Calafut:2021wkx}. It would be interesting to consider how the presence of residual tSZ or CIB foreground components, or emission from radio and dusty star-forming galaxies, might affect, or be identified through, the machine learning approach. We leave the integration of such extensions to future work.

In summary, this work provides a promising method to measure the pairwise peculiar velocity estimator using a machine learning model. The model utilizes the relationships between cluster observables derived from high-fidelity simulations to predict the velocity directly from the kSZ data rather than measuring additional optical depth information derived from observations. The approach will be valuable in analyses of the upcoming CMB experiments, from the CCAT, Simons Observatory, and CMB-S4 surveys, along with galaxy surveys from DESI, SDSS, Euclid, and LSST to use cluster velocity correlations to constrain the properties of gravity, dark energy and neutrinos.

\begin{acknowledgments}

We thank Patricio Gallardo for making available the kSZ pairwise correlation code from \citep{Calafut:2021wkx} that is used in this work. We thank Nicholas Battaglia and Eve Vavagiakis for their helpful discussions and comments on the manuscript.

The work of YG and RB is supported by NSF grant AST-2206088, NASA ATP grant 80NSSC18K0695, and NASA ROSES grant 12-EUCLID12-0004.

\end{acknowledgments}

\bibliographystyle{apsrev}
\bibliography{ML}

\end{document}